\documentclass{appolb}
\usepackage{graphicx}
\usepackage{braket}

\usepackage{cite}

\newcommand{\Pom}{\mathbb{P}}
\newcommand{\Reg}{\mathbb{R}}
\usepackage{amsmath,amsfonts,amssymb,amstext,mathrsfs}

\begin{document}
\title{The Tensor Pomeron and Low-x Deeply Virtual Compton Scattering
\thanks{Presented by P. Lebiedowicz at XXIX Cracow Epiphany Conference
on Physics at the Electron-Ion Collider and Future Facilities,
Krak{\'o}w, Poland,
January 16-19, 2023.}%
}
\author{Piotr Lebiedowicz, Antoni Szczurek
\address{Institute of Nuclear Physics Polish Academy of Sciences, 
Radzikowskiego 152, PL-31342 Krak{\'o}w, Poland}
\\[3mm]
{Otto Nachtmann 
\address{Institut f\"ur Theoretische Physik, Universit\"at Heidelberg,
Philosophenweg 16, D-69120 Heidelberg, Germany}
}
\\[3mm]
}
\maketitle
\begin{abstract}
The two-tensor-pomeron model is applied to 
deeply virtual Compton scattering (DVCS) on a proton.
A good description of the DVCS HERA data 
at small Bjorken-$x$ 
is achieved due to a sizeable interference 
of soft and hard pomeron contributions.
We present two fits which differ somewhat in the strength
of the hard pomeron contribution.
We describe, in the same framework,
both the low $Q^{2}$ and high $Q^{2}$ regimes 
and the transition between them.
We find that the soft-pomeron contribution is considerable 
up to $Q^{2} \sim 20$~GeV$^{2}$.
The reggeon exchange term is particularly
relevant for describing 
the scattering of a real photon on a proton 
measured at lower $\gamma p$ energies at FNAL.
We find that the ratio of $\gamma^{*} p \to \gamma p$ cross-sections
for longitudinally and transversely polarized
virtual photons strongly increases with $t$.
Our findings may be checked
in future lepton-nucleon scattering experiments in the low-$x$ regime,
for instance, at a future Electron-Ion Collider (EIC) at the BNL
and LHeC at the LHC.
\end{abstract}
  
\section{Introduction}
\label{sec:introduction}

This presentation summarizes some of the
key results of \cite{Lebiedowicz:2022xgi} 
to which we refer the reader for further details.
We apply the two-tensor-pomeron model \cite{Britzger:2019lvc}
to deeply virtual Compton scattering (DVCS) on the proton,
$\gamma^{*} p \to \gamma p$.
Our model can be used
for large $\gamma^{*} p$ c.m. energy 
$W \gg m_{p}, \sqrt{|t|}$, $|t| \lesssim 1$~GeV$^{2}$,
and small Bjorken-$x$, say
$x = Q^{2}/(W^{2} + Q^{2} - m_{p}^{2}) < 0.02$.
Here $m_{p}$ is the proton mass,
$Q^{2}$ is the photon virtuality, and
$t$ is the squared momentum transfer at the proton vertex.

The DVCS has been a subject of extensive experimental
and theoretical research; 
see Sec.~1 in \cite{Lebiedowicz:2022xgi}.
Experimental program at the Electron-Ion Collider (EIC)
\cite{AbdulKhalek:2021gbh},
and, if realized, 
at the Large Hadron Electron Collider (LHeC) 
at the LHC \cite{LHeC:2020van},
are expected to improve our knowledge of DVCS
in a wide kinematic range.

DVCS is a prime playground for the application
of the generalized parton-distribution (GPD) concept
based on perturbative QCD (pQCD), 
cf.~\cite{Diehl:2003ny} for a review.
Here we discuss DVCS in the Regge approach where
the scattering is described using exchange objects.
In the tensor-pomeron model, 
introduced for soft high-energy reactions in \cite{Ewerz:2013kda}
and extended to hard reactions in \cite{Britzger:2019lvc},
the charge-conjugation
$C = +1$ soft ($\Pom_{1}$) and hard ($\Pom_{0}$) pomeron 
and the reggeons ($\Reg_{+} = f_{2 \Reg}, a_{2 \Reg}$)
are described as effective rank-2 symmetric tensor exchanges.

A two-pomeron description of low-$x$ DIS 
was first proposed in \cite{Donnachie:1998gm}.
However, there a vector nature of the pomerons was considered.
It was shown in \cite{Britzger:2019lvc} 
that considering these two pomerons as vector objects leads
to the conclusion that they decouple in the total photoabsorption
cross-section on the proton and 
in the structure functions of low-$x$ DIS.
But experiment clearly shows pomeron-exchange behaviour
for these quantities at large $W$;
see Fig.~5 of \cite{Britzger:2019lvc}.

Applications of the tensor-pomeron approach
were given for a number of exclusive central-production reactions,
see e.g. \cite{Lebiedowicz:2013ika,Lebiedowicz:2014bea,Lebiedowicz:2016ioh,Lebiedowicz:2018sdt,Lebiedowicz:2019jru,Lebiedowicz:2019boz,Lebiedowicz:2020yre},
and for soft-photon radiation 
\cite{Lebiedowicz:2021byo,Lebiedowicz:2022nnn,Lebiedowicz:2023mhe}
in hadron-hadron collisions.
For some remarks on the history of tensor-pomeron concepts
and corresponding references see \cite{Ewerz:2016onn}.
It is worth mentioning that
the tensor-pomeron current (Eq.~(2.3) of \cite{Ewerz:2016onn})
cannot be universally proportional to the energy-momentum tensor.

\section{Formalism}
\label{sec:formalism}

\begin{figure}[!h]
\includegraphics[width=5cm]{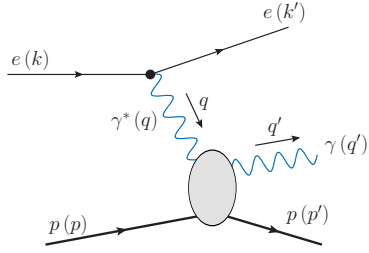}
\caption{DVCS contribution to $ep \to ep \gamma$ (\ref{2.100}).}
\label{fig:1}
\end{figure}

We investigate the real and deeply virtual Compton scattering 
on a proton
\begin{equation}
\gamma^{(*)}(q, \epsilon) + p (p,\lambda) \to 
\gamma(q', \epsilon') + p (p',\lambda')\,.
\label{2.1}
\end{equation}
The momenta are indicated in brackets,
$\lambda, \lambda' \in \{1/2, -1/2 \}$
are the proton helicities,
and $\epsilon, \epsilon'$ are
the photon polarization vectors.
For an initial virtual photon $\gamma^{*}$ the reaction (\ref{2.1})
is extracted from $ep \to ep \gamma$ scattering
(see Fig.~\ref{fig:1})
\begin{equation}
e(k) + p (p,\lambda) \to 
e(k') + \gamma(q', \epsilon') + p (p',\lambda')\,.
\label{2.100}
\end{equation}
Here the Bethe-Heitler process and DVCS contribute with
the latter corresponding to electroproduction of the $\gamma p$ state.
We assume for (\ref{2.100}) unpolarized initial particles
and no observation of the polarization of final state particles.
The standard kinematic variables
are (see Table~1 of \cite{Arens:1996xw})
\begin{eqnarray}
&& q = k - k'\,, \; q^{2} = -Q^{2}\,, \; 
s = (p + k)^{2}\,, \;
t = (p - p')^{2}\,, \;
W^{2} = (p + q)^{2}\,, \nonumber \\
&& x = \frac{Q^{2}}{2 p \cdot q}
     = \frac{Q^{2}}{W^{2} + Q^{2} - m_{p}^{2}} \,,
     \;
y = \frac{p \cdot q}{p \cdot k}
     = \frac{W^{2} + Q^{2} - m_{p}^{2}}{s - m_{p}^{2}}\,.
\label{2.101}
\end{eqnarray}
Adapting (3.20) and (3.21) of \cite{Arens:1996xw} 
to the reaction (\ref{2.100})
and integrating over the azimuthal angle $\varphi$
defined in (2.1) of \cite{Arens:1996xw}
we get for the DVCS part:
\begin{eqnarray}
\frac{d\sigma (ep \to ep \gamma)}{dydQ^{2}dt}
= \Gamma_{e p \gamma} 
\left( 
\frac{d\sigma_{\rm T}}{dt}(Q^{2},W^{2},t) 
+ \varepsilon \frac{d\sigma_{\rm L}}{dt}(Q^{2},W^{2},t)
\right)
\label{2.102}
\end{eqnarray}
with $\Gamma_{e p \gamma}$ the $\gamma^{*}$ flux factor \cite{Lebiedowicz:2022xgi}.
The differential cross-sections
for $\gamma^{*} p \to \gamma p$ for transverse and longitudinal
polarization of the~$\gamma^{*}$ are
\begin{eqnarray}
&&\frac{d\sigma_{\rm T}}{dt}(Q^{2},W^{2},t) =
\frac{1}{2} \left( \frac{d\sigma_{++}}{dt}(Q^{2},W^{2},t) 
                 + \frac{d\sigma_{--}}{dt}(Q^{2},W^{2},t) \right)
                 \,, \nonumber\\
&&\frac{d\sigma_{\rm L}}{dt}(Q^{2},W^{2},t) = 
\frac{d\sigma_{00}}{dt}(Q^{2},W^{2},t)\,,
\label{2.104}
\end{eqnarray}
where
\begin{eqnarray}
&&\frac{d\sigma_{mm}}{dt}(Q^{2},W^{2},t)
=\frac{1}{16 \pi (W^{2} - m_{p}^{2}) 
\sqrt{(W^{2} - m_{p}^{2} + Q^{2})^{2} + 4 m_{p}^{2} Q^{2}}} \nonumber\\
&& \qquad \qquad \times 
\frac{1}{2}\sum_{\lambda, \lambda', a}
\big\vert \Braket{\gamma(q', \epsilon'_{a}),p (p',\lambda'), {\rm out}
|e J_{\mu}(0) \epsilon_{m}^{\mu}| p (p,\lambda)} \big\vert^{2} \,.
\label{2.105}
\end{eqnarray}
Here $e J_{\mu}$ is the electromagnetic current,
$\epsilon_{m} (m = \pm, 0)$ are the standard $\gamma^{*}$
polarization vectors for right and left circular
and longitudinal polarization
(see (3.11)--(3.14) of \cite{Arens:1996xw})
and $\epsilon'_{a} (a = 1, 2)$
are the polarization vectors of the real photon in the final state.

We describe the amplitude for (\ref{2.1})
in terms of the $C = +1$ exchanges of soft ($\Pom_{1}$)
and hard ($\Pom_{0}$) pomeron and the reggeons
$f_{2 \Reg}$ and $a_{2 \Reg}$.
We get
\begin{eqnarray}
&&\Braket{
\gamma(q', \epsilon'),p (p',\lambda'), {\rm out}
|e J_{\nu}(0) \epsilon^{\nu}| p (p,\lambda)} 
\equiv
(\epsilon'^{\mu})^* {\cal M}_{\mu \nu, \lambda' \lambda}\, \epsilon^{\nu}
\nonumber \\
&& 
=-(\epsilon'^{\mu})^*  
\sum_{j = 0, 1}
\Gamma_{\mu \nu \kappa \rho}^{(\Pom_{j} \gamma^{*} \gamma^{*})}(q',q)\, \epsilon^{\nu}\,
\Delta^{(\Pom_{j})\,\kappa \rho, \alpha\beta}(W^{2},t)\, 
\bar{u}_{\lambda'}(p') 
\Gamma_{\alpha\beta}^{(\Pom_{j} pp)}(p',p)
u_{\lambda}(p) 
\nonumber \\  
&& \quad + (\Pom_{j} \to f_{2 \Reg}, a_{2 \Reg})\,.
\label{2.3}
\end{eqnarray}
Here $\Delta^{(\Pom_{j})}$, $\Gamma^{(\Pom_{j} pp)}$, and
$\Gamma^{(\Pom_{j} \gamma^{*} \gamma^{*})}_{\mu \nu \kappa \rho}(q',q)$
denote the effective propagator, the proton vertex function,
and the $\Pom_{j} \gamma^{*} \gamma^{*}$ ($j = 0,1$) vertex,
respectively, for the tensor pomerons $\Pom_{j}$.
The \textit{Ans{\"a}tze} 
for the $f_{2 \Reg}$ and $a_{2 \Reg}$ reggeons
($j = 2$) are analogous.
The properties of $\Pom_{0}, \Pom_{1}, f_{2 \Reg}$, 
and $a_{2 \Reg}$, and their couplings to protons and photons
will be taken as in \cite{Britzger:2019lvc}.
We get
\begin{eqnarray}
&&{\cal M}_{\mu \nu, \lambda' \lambda}
=-i \sum_{j = 0, 1, 2}
\frac{3 \beta_{jpp} }{2W^{2}} 
(-i W^{2} \alpha'_{j})^{\alpha_{j}(t)-1}\,
F_{1}^{(j)}(t) 
\bar{u}_{\lambda'}(p') 
\gamma^{\kappa}(p' + p)^{\rho}
u_{\lambda}(p) \nonumber \\
&&\qquad \times
\left[ 2a_{j \gamma^{*} \gamma^{*}}(q^{2},0,t)\,\Gamma^{(0)}_{\mu \nu \kappa \rho}(q',-q)\,
- b_{j \gamma^{*} \gamma^{*}}(q^{2},0,t)\,\Gamma^{(2)}_{\mu \nu \kappa \rho}(q',-q) \right]
 \,.
\label{2.200}
\end{eqnarray}
For the coupling constants $\beta_{j pp}$ 
of the pomerons $(j=0,1)$ and reegeon $(j=2)$ to protons we take
$\beta_{0 pp} = \beta_{1 pp} = 1.87$~GeV$^{-1}$,
$\beta_{2 pp} = 3.68$~GeV$^{-1}$.
The pomeron and reggeon trajectory functions
are assumed to be of linear form
\begin{eqnarray}
\alpha_{j}(t) = \alpha_{j}(0)+\alpha'_{j}\,t\,,
\quad
\alpha_{j}(0) = 1 + \epsilon_{j}\,,
\quad
(j = 0, 1, 2)\,.
\label{trajectories}
\end{eqnarray}
The slope parameters $\alpha'_{j}$ 
are taken as the default values from 
\cite{Britzger:2019lvc}:
$\alpha'_{1} = \alpha'_{0} = 0.25\;{\rm GeV}^{-2}$,
$\alpha'_{2} = 0.9\;{\rm GeV}^{-2}$.
The intercept parameters of the trajectories (\ref{trajectories})
were determined from detailed comparison
of the model with the DIS HERA data
and photoproduction data in~\cite{Britzger:2019lvc}:
\begin{eqnarray}
{\rm soft \; pomeron} \;\Pom_{1}:\quad &&\epsilon_{1} = 0.0935(^{+76}_{-64})\,,
\label{FIT_parameters_P1}
\\
{\rm hard \; pomeron} \;\Pom_{0}:\quad &&\epsilon_{0} = 0.3008(^{+73}_{-84})\,,
\label{FIT_parameters_P0}\\
{\rm reggeon} \;\Reg_{+}:\quad &&\alpha_{2}(0) = 0.485(^{+88}_{-90})\,.
\label{FIT_parameters_R}
\end{eqnarray}
The coupling functions $a_{j \gamma^{*} \gamma^{*}}$ 
and $b_{j \gamma^{*} \gamma^{*}}$ in (\ref{2.200})
are (see (2.21)--(2.23) of \cite{Lebiedowicz:2022xgi})
\begin{eqnarray}
&&a_{j \gamma^{*} \gamma^{*}}(q^{2},0,t) = 
e^{2} \hat{a}_{j}(Q^{2}) F^{(j)}(t)\,, \quad j = 0, 1, 2\,,
\nonumber \\
&&b_{2 \gamma^{*} \gamma^{*}}(q^{2},0,t) = 
e^{2} \hat{b}_{2}(Q^{2}) F^{(2)}(t)\,.
\label{A4}
\end{eqnarray}  
Here,
$\hat{a}_{j}(Q^{2})$ and $\hat{b}_{j}(Q^{2})$ 
were determined in \cite{Britzger:2019lvc}
from the global fit to HERA inclusive DIS data
for $Q^{2} < 50$~GeV$^{2}$ and $x < 0.01$
and the ($Q^{2} = 0$) photoproduction data.
All coupling functions $\hat{a}_{j}$ and $\hat{b}_{j}$ 
are plotted in Fig.~2 of \cite{Lebiedowicz:2022xgi}.
For small $Q^{2}$, the soft pomeron function $b_{1 \gamma^{*} \gamma^{*}}$
gives a larger contribution to the cross-section
than the corresponding hard one $b_{0 \gamma^{*} \gamma^{*}}$.
In the large $Q^{2}$ region the reverse is found.
In Sec.~\ref{sec:results}, we shall show
two alternative fits for $b_{1 \gamma^{*} \gamma^{*}}$
and $b_{0 \gamma^{*} \gamma^{*}}$
obtained from a comparison to HERA DVCS data.

We use the combined form-factor functions
for a given $j$~$(j = 0, 1, 2)$
\begin{equation}
F_{\rm eff}^{(j)}(t) = 
F^{(j)}(t) \times  F_{1}^{(j)}(t) =
\exp(-b_{j}|t|/2 )\,,
\label{t_dependence_ff}
\end{equation}
assuming the same $t$ dependence for both $a$ and $b$ 
coupling functions.
We take
$b_{1} = b_{2} = 5.0\; {\rm GeV}^{-2}$
and 
$b_{0} = 1.0\; {\rm GeV}^{-2}$ from \cite{Lebiedowicz:2022xgi}.

\section{Results}
\label{sec:results}
We shall restrict our discussion to experimental results
that satisfy the conditions $x \approx Q^{2}/W^{2} < 0.02$
and $|t| \lesssim 1$~GeV$^{2}$ where our model should be reliable.

\begin{figure}[htb]
\includegraphics[width=0.42\textwidth]{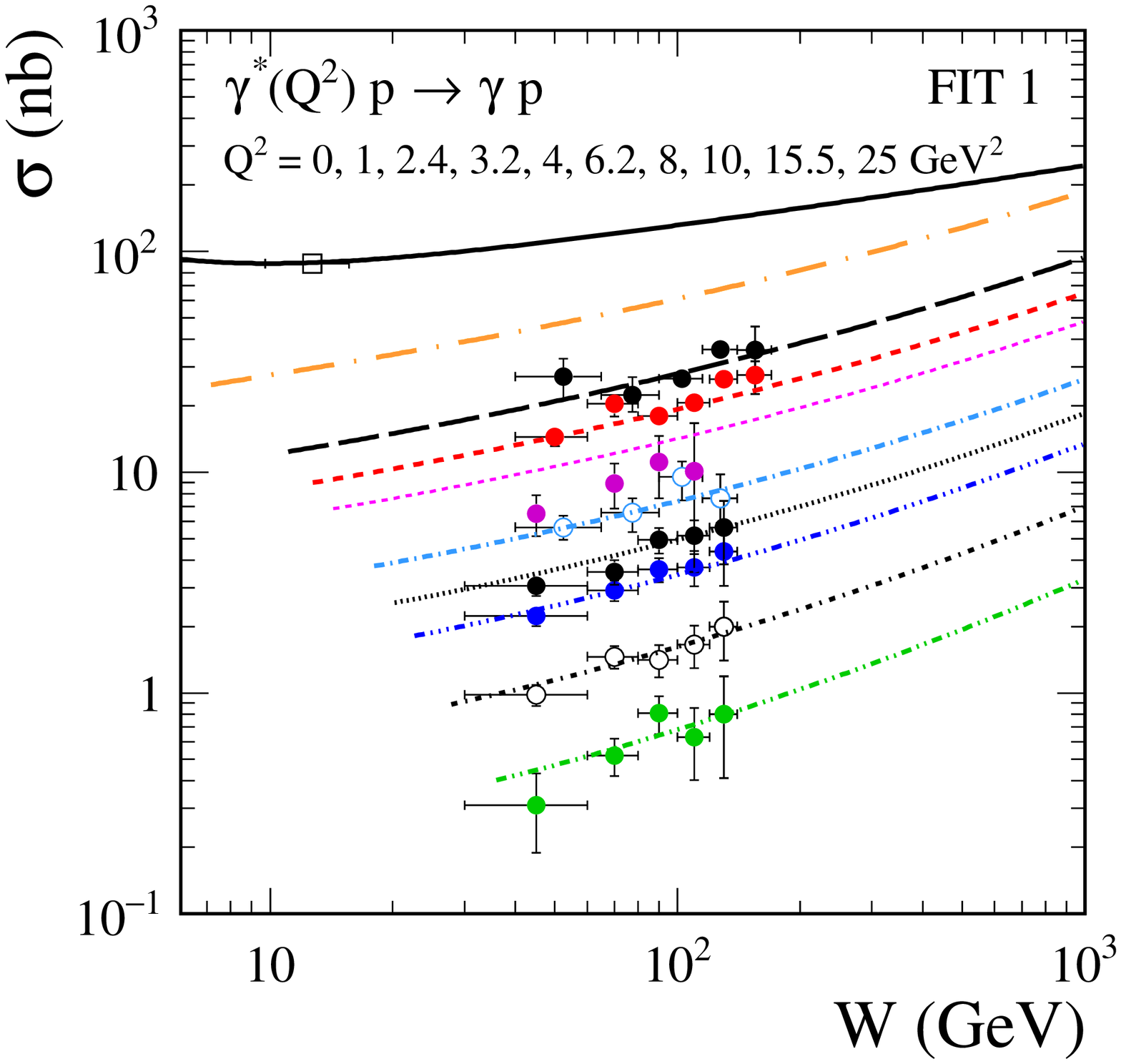}
\includegraphics[width=0.42\textwidth]{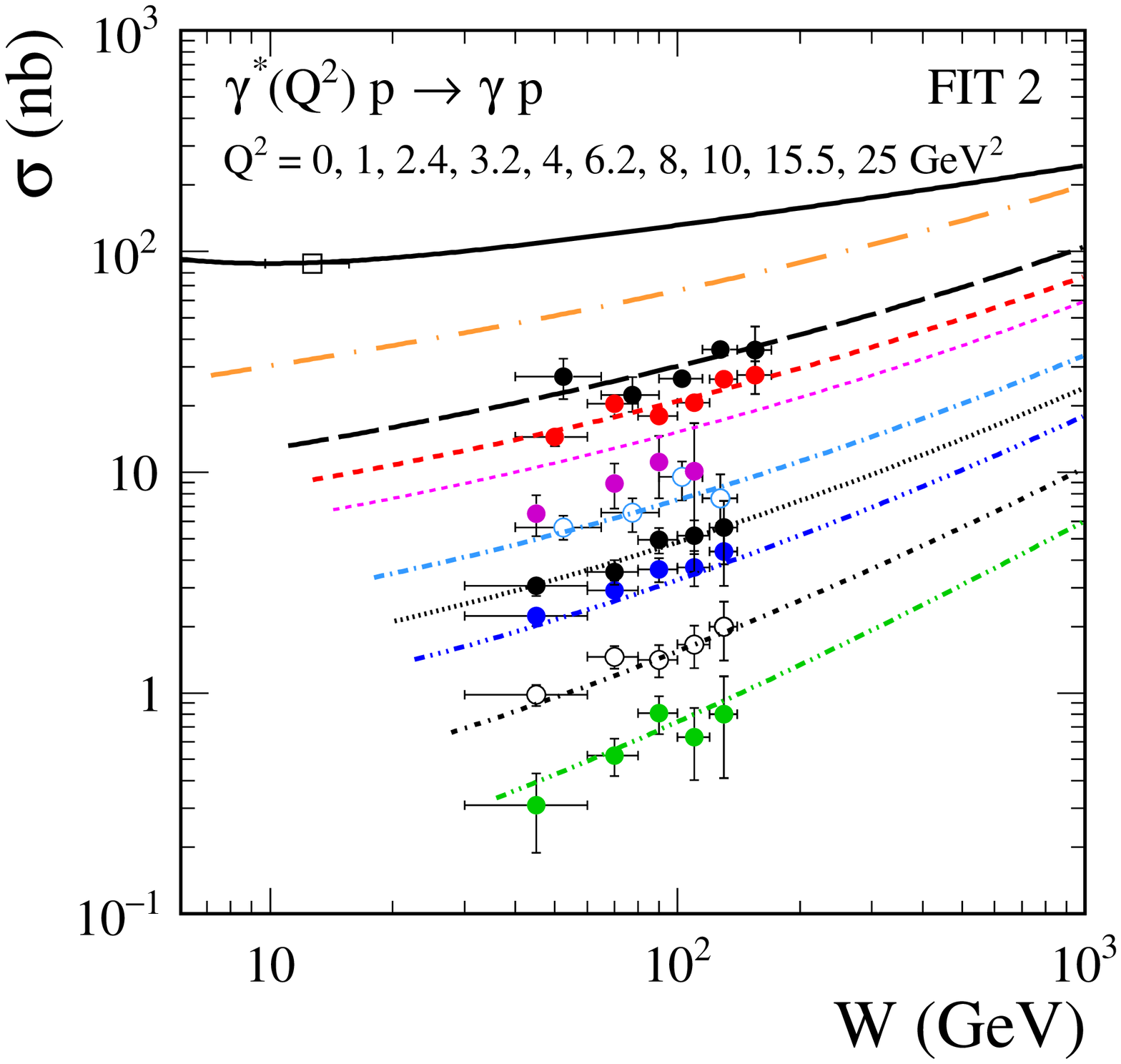}\\
\includegraphics[width=0.42\textwidth]{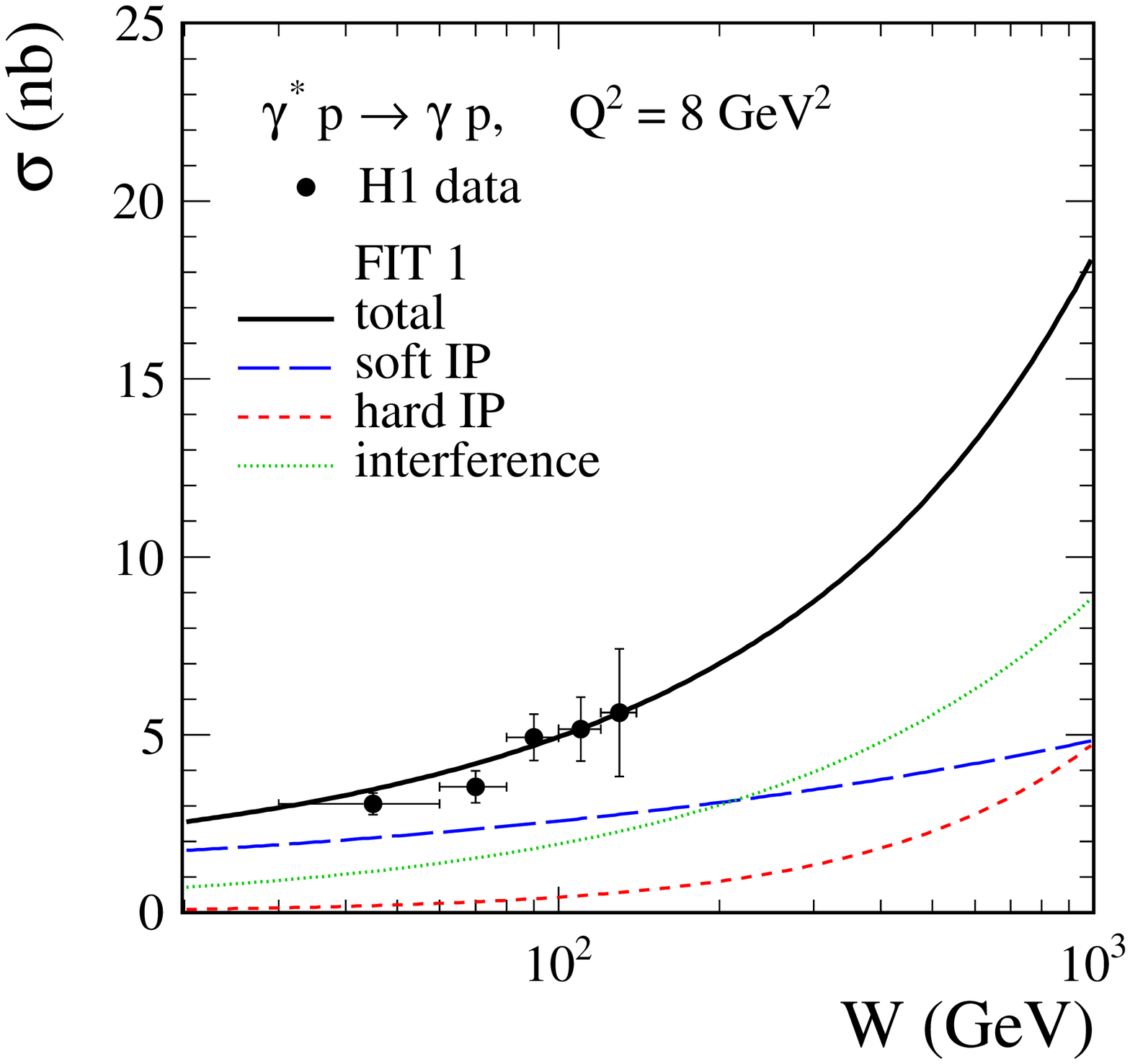}
\includegraphics[width=0.42\textwidth]{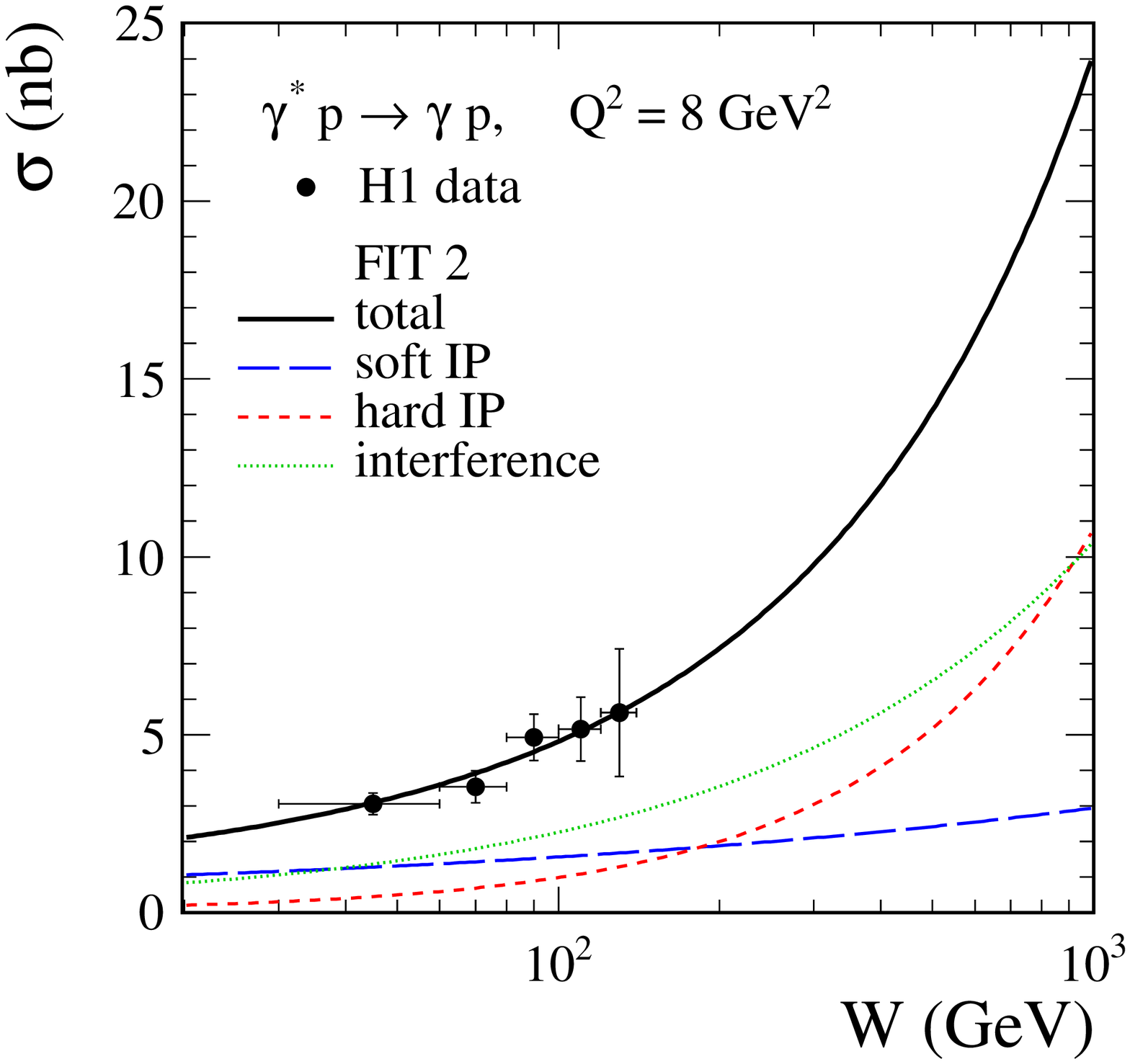}\\
\includegraphics[width=0.42\textwidth]{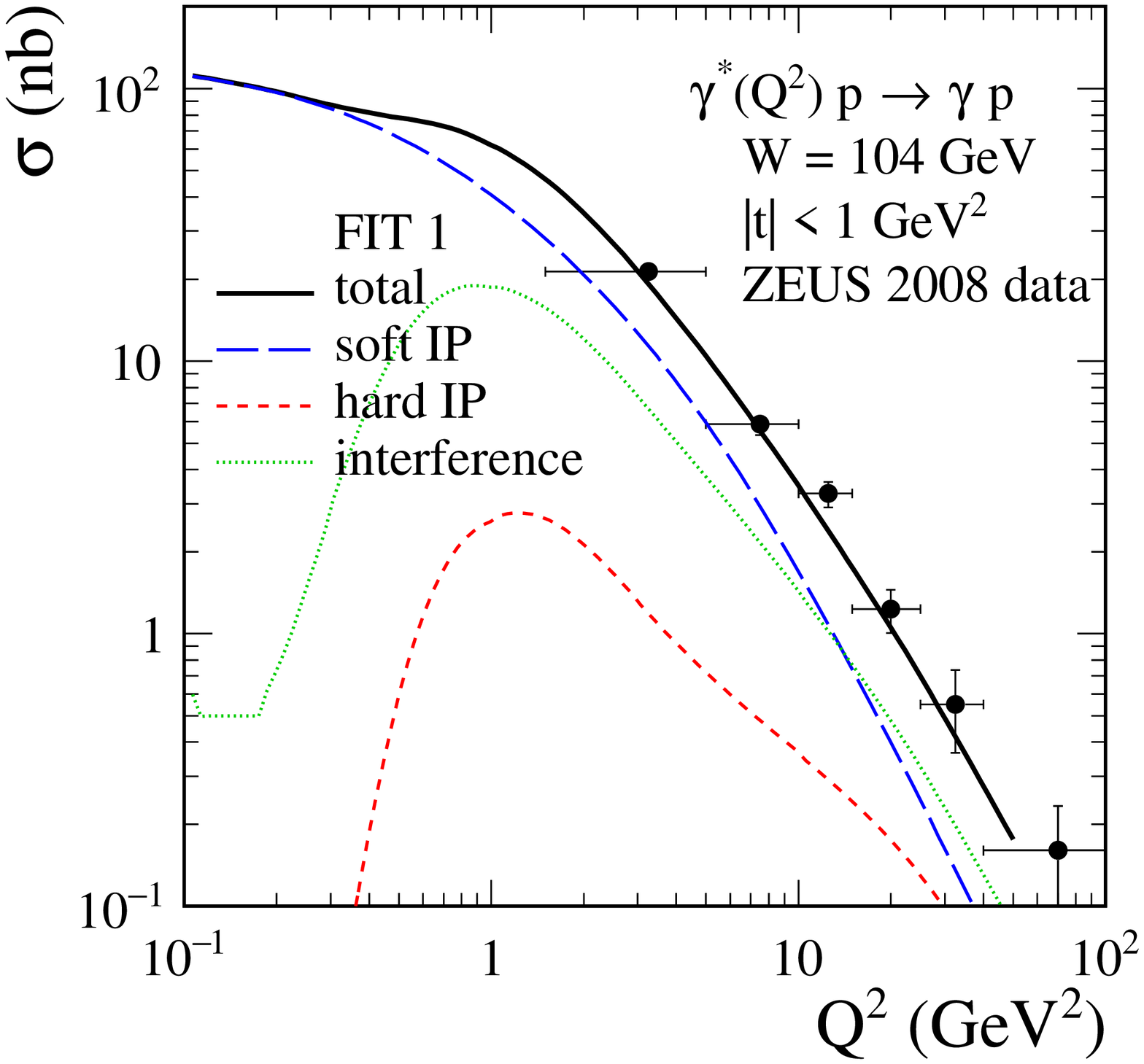}
\includegraphics[width=0.42\textwidth]{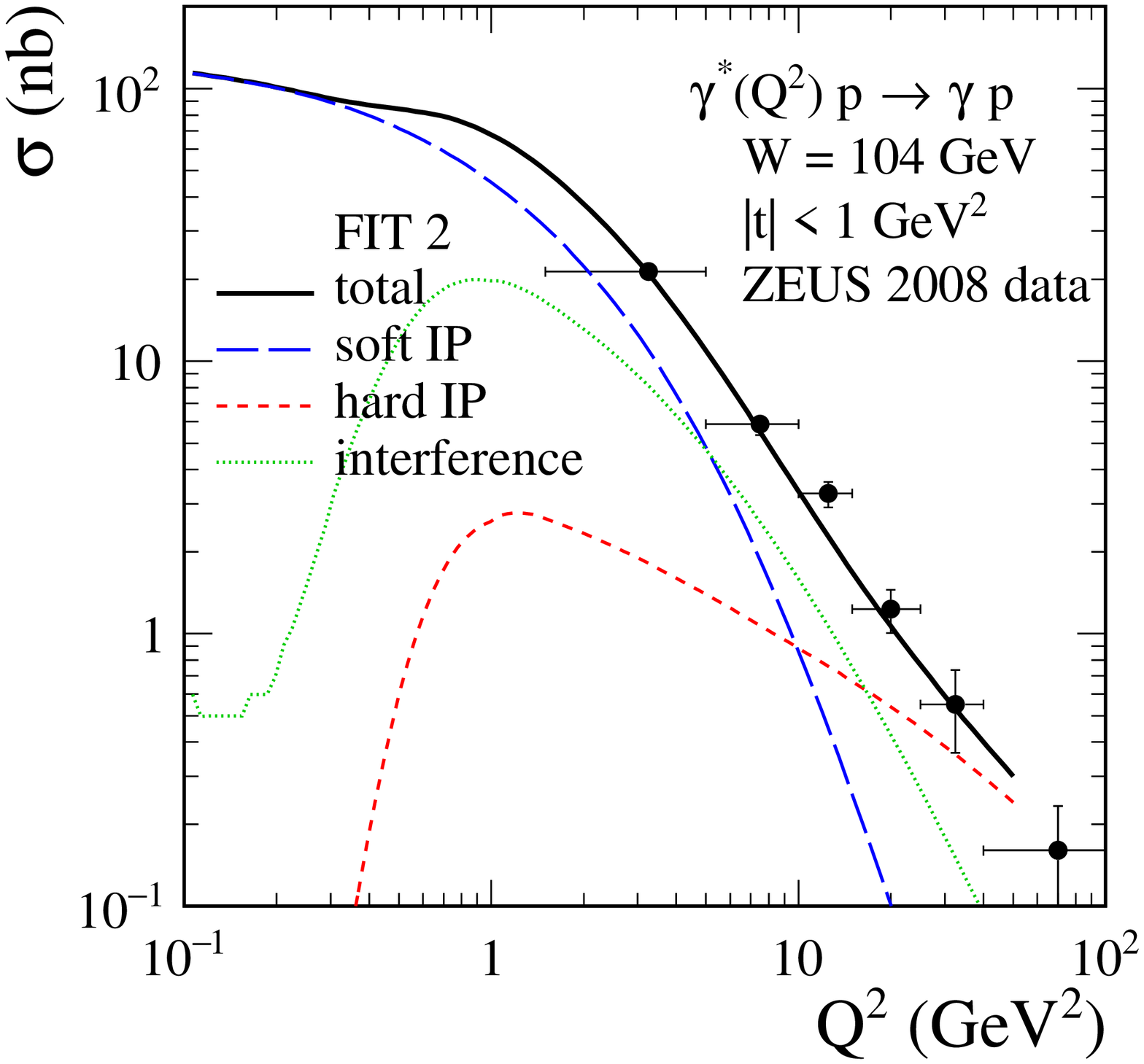}
\caption{\textbf{Top panels}: 
Total cross-sections as a function of the c.m. energy $W$
for FIT~1 (left) and FIT~2 (right).
Comparison of theoretical results
to the FNAL data from \cite{Breakstone:1981wm} 
for real Compton scattering ($Q^{2} = 0$)
and to the DVCS HERA data is shown.
The upper black solid line is for $Q^{2} = 0$,
the orange long-dashed-dotted line is for $Q^{2} = 1$~GeV$^{2}$.
The remaining lines correspond to the values
\mbox{$Q^{2} = 2.4, 3.2, 4,
 6.2, 8, 10, 15.5, 25$~GeV$^{2}$}
(from top to bottom) and should be compared with 
the HERA data from
\cite{ZEUS:2003pwh,H1:2005gdw,ZEUS:2008hcd,H1:2009wnw}.
\textbf{Middle panels}:
Fit results for $Q^{2} = 8$~GeV$^{2}$
together with the H1 data \cite{H1:2009wnw}.
The interference term of soft and hard pomeron
is shown by the green dotted line.
\textbf{Bottom panels}:
Comparison of cross-sections
as a function of $Q^{2}$ to the ZEUS data from \cite{ZEUS:2008hcd}.}
\label{fig:2}
\end{figure}

In Fig.~\ref{fig:2} we compare
the tensor-pomeron model results,
FIT~1 (left panel) and FIT~2 (right panel),
to the FNAL data 
\cite{Breakstone:1981wm}
on real-photon-proton scattering ($\gamma p \to \gamma p$),
and to HERA data
\cite{ZEUS:2003pwh,H1:2005gdw,ZEUS:2008hcd,H1:2009wnw} 
on DVCS ($\gamma^{*}(Q^{2}) p \to \gamma p$)
for different averaged $W$ and $Q^{2}$.
The complete cross-section is a coherent sum of
soft and hard components in the amplitude.
For real Compton scattering ($Q^{2} = 0$)
the cross-section is dominated by soft-pomeron exchange with
an additional contribution from reggeon exchange at lower energies.
The hard-pomeron contribution is negligibly small there.
The dominant contribution 
comes from the $b$-type coupling functions
$b_{j \gamma^{*}\gamma^{*}}$ ($j =0,1$).
Their size differs in the two fits.
We see from the bottom panels of Fig.~\ref{fig:2} that
for higher $Q^{2}$ the soft component slowly decreases
relative to the hard one.
A significant constructive interference effect 
between the soft and hard components is clearly visible.
Here and in the following, the interference term
is calculated as the difference of 
coherent and incoherent cross-sections
of the $\Pom_{1}$, $\Pom_{0}$, and $\Reg$ contributions.
The Fits~1 and 2 hardly differ for the $W$ region where
there are data. But for higher $W$ values FIT~2, 
where the contribution from the hard pomeron is enhanced,
gives a steeper rise of the cross-section with $W$ 
and especially so for larger $Q^{2}$.
For FIT~1, we see that
the soft pomeron survives to relatively large $Q^2$
and at $Q^2 \simeq 50$~GeV$^{2}$ the interference term
plays an important role in the description of the data.

\begin{figure}[htb]
\includegraphics[width=0.42\textwidth]{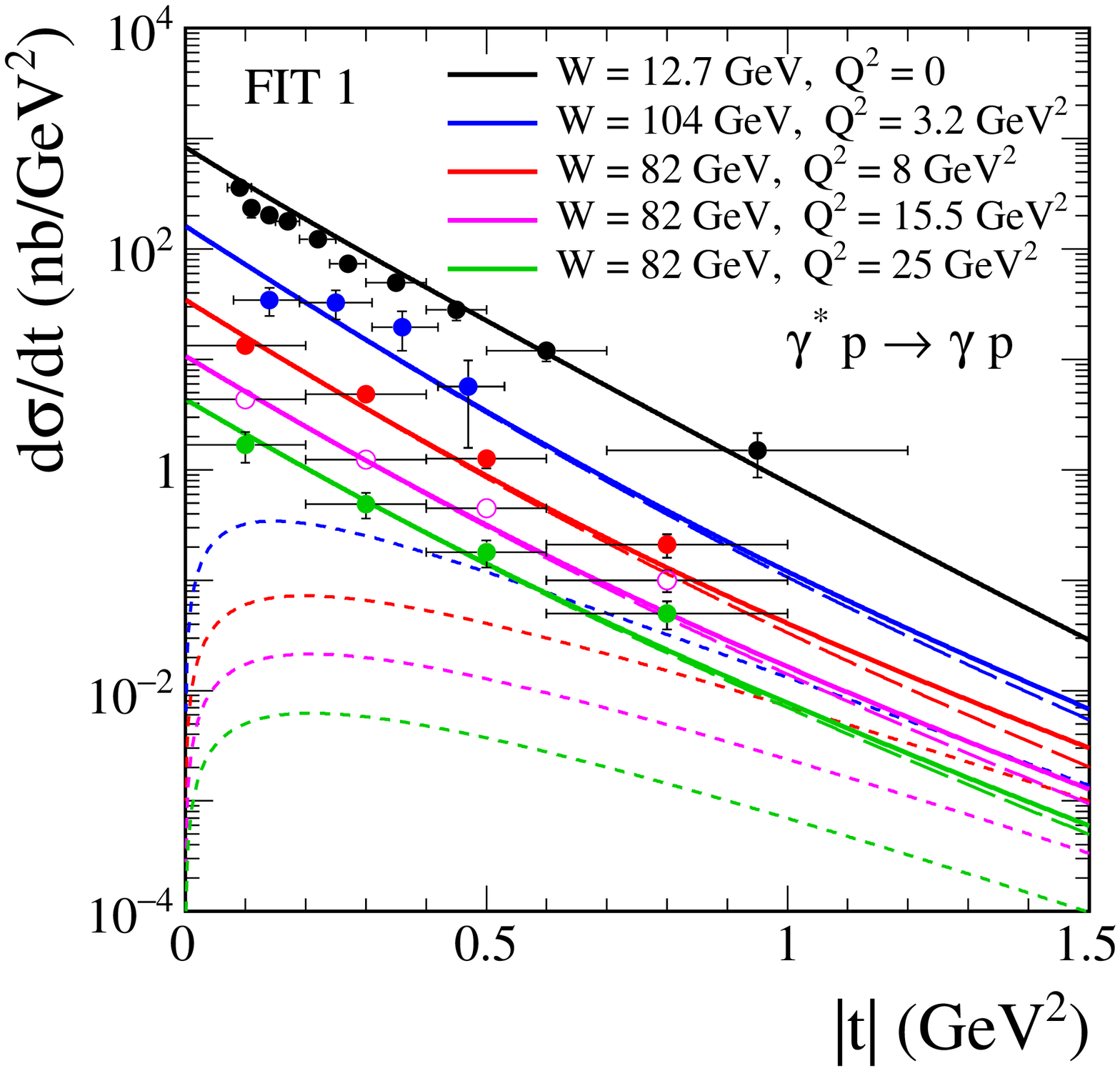}
\includegraphics[width=0.42\textwidth]{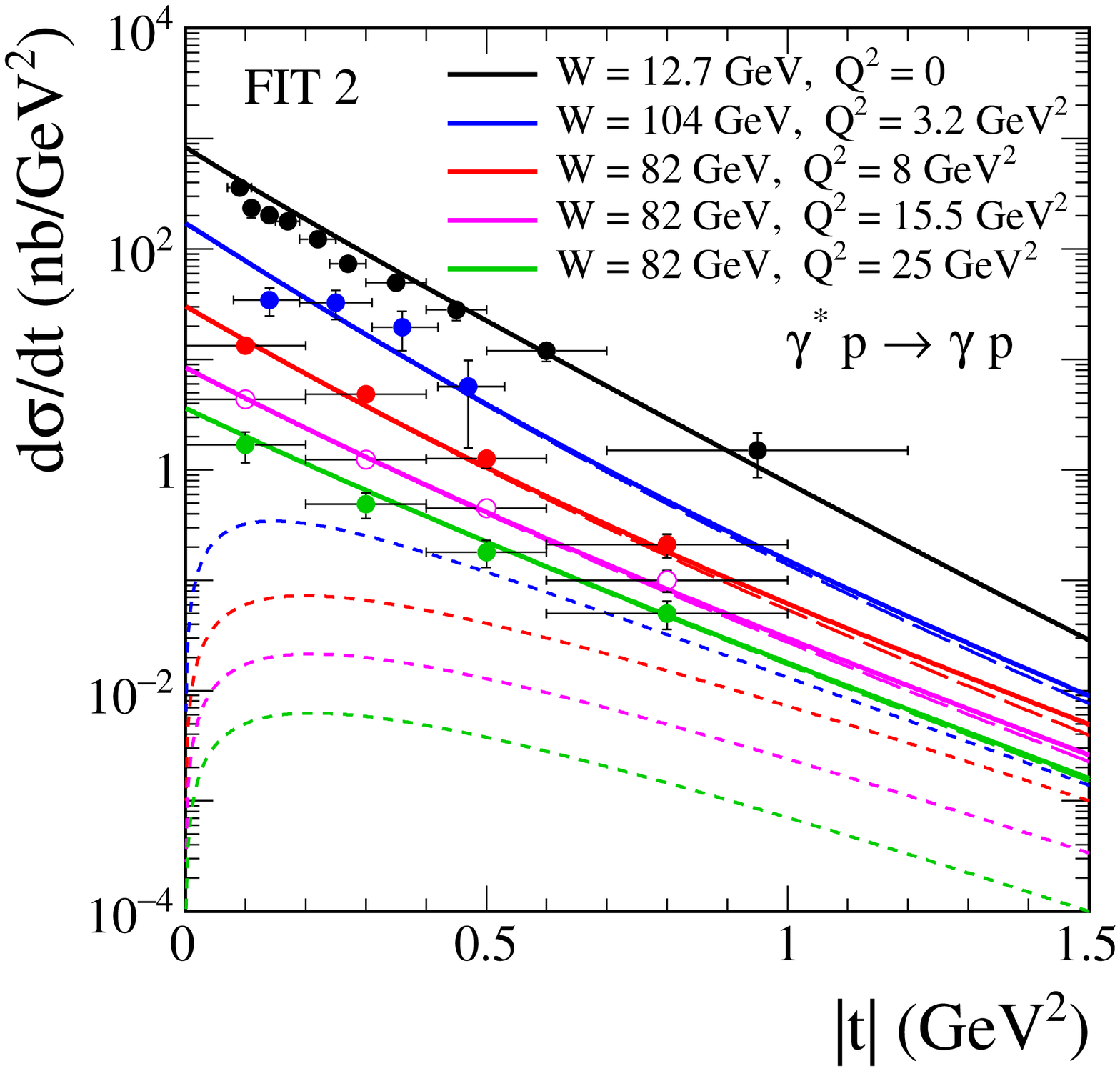}\\
\includegraphics[width=0.42\textwidth]{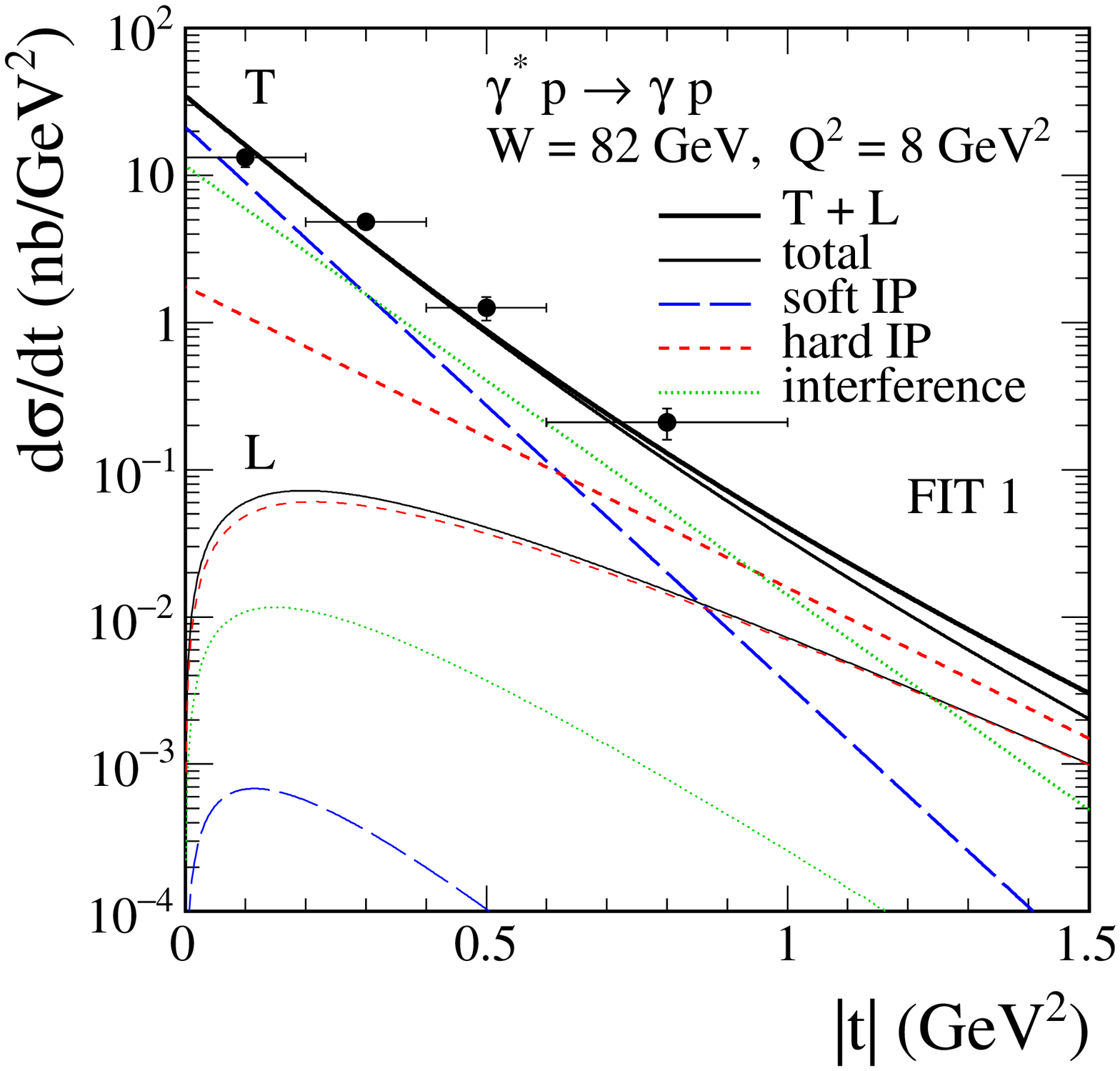}
\includegraphics[width=0.42\textwidth]{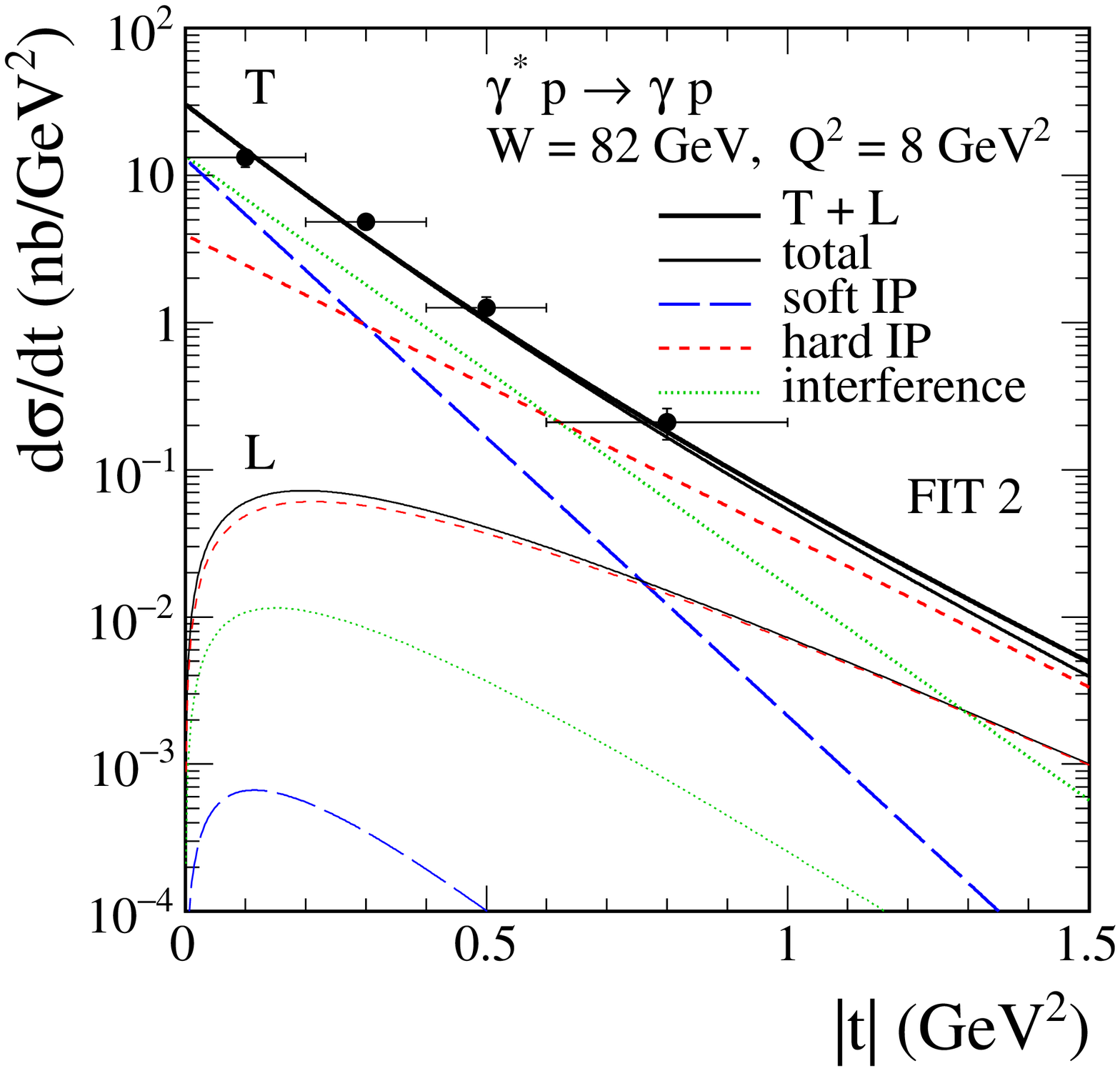}\\
\includegraphics[width=0.42\textwidth]{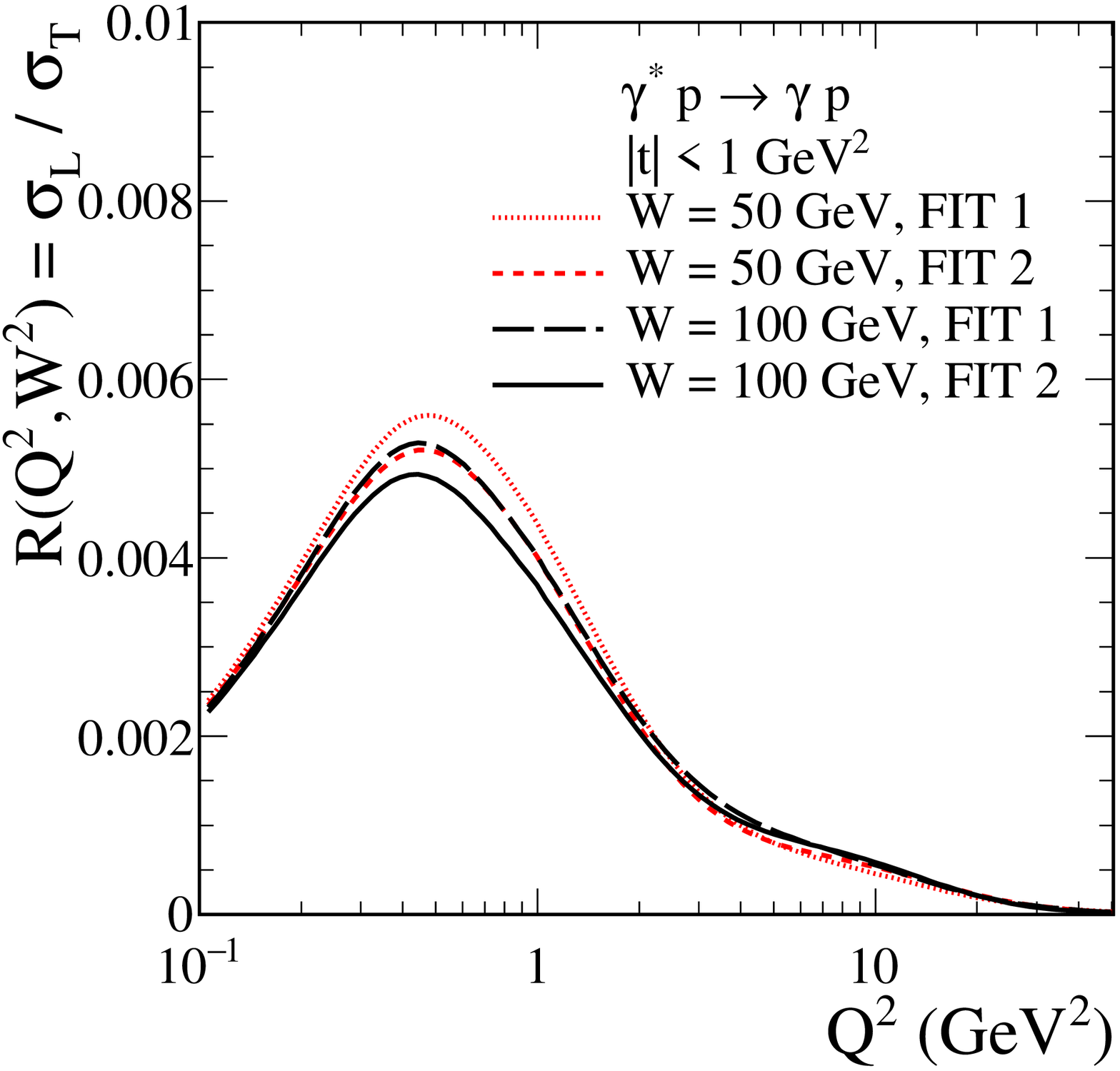}
\includegraphics[width=0.42\textwidth]{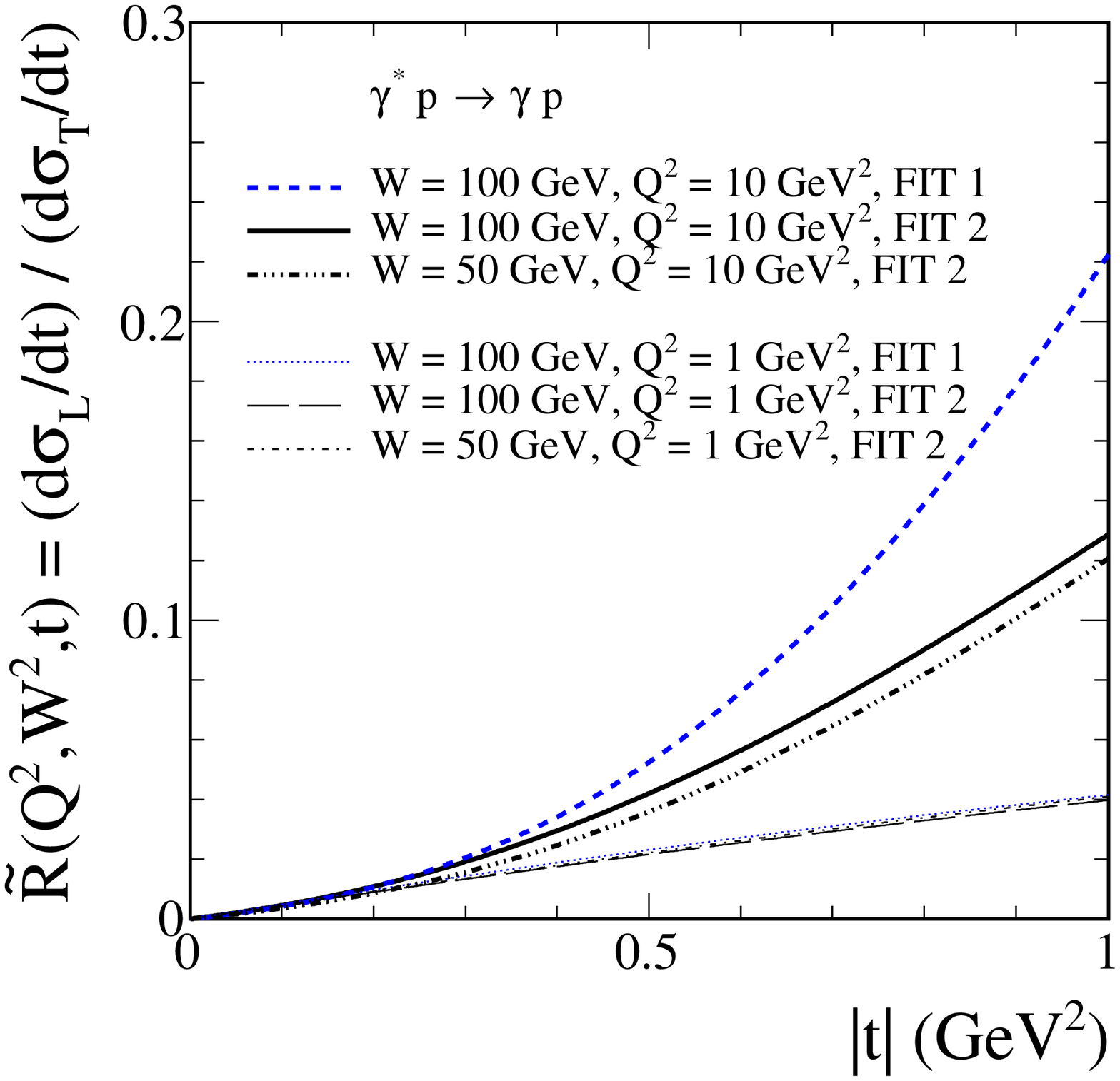}
\caption{\textbf{Top panels}: 
The differential cross-sections $d\sigma/dt$
for FIT~1 (left) and FIT~2 (right)
compared to experimental data for different $W$ and $Q^{2}$.
The upper line corresponds to $W = 12.7$~GeV 
and $Q^{2} = 0$ and averaged FNAL data \cite{Breakstone:1981wm}.
The bottom lines correspond to 
theoretical results for $\gamma^{*} p \to \gamma p$ at higher $Q^{2}$ 
for transverse (long-dashed lines) and longitudinal (short-dashed lines)
polarization of the $\gamma^{*}$ and their sum (solid lines).
Data for $\langle W \rangle = 104$~GeV are from \cite{ZEUS:2008hcd}
and for $\langle W \rangle = 82$~GeV from \cite{H1:2009wnw}.
\textbf{Middle panels}: 
Comparison of the results for $\gamma^{*} p \to \gamma p$ 
for transverse (${\rm T}$) and longitudinal (${\rm L}$)
polarization of the $\gamma^{*}$ individually 
and their sum ${\rm T + L}$ (see the upper solid lines) 
to DVCS H1 data \cite{H1:2009wnw}
for $W = 82$~GeV and $Q^{2} = 8$~GeV$^{2}$.
\textbf{Bottom panels}: The ratios (\ref{ratio_t})
for the $\gamma^{*} p \to \gamma p$ reaction.
Note that the meaning of the lines on these two panels is different.}
\label{fig:3}
\end{figure}

In Fig.~\ref{fig:3} we show the differential cross-sections 
$d\sigma/dt$
for different $\langle W \rangle$ and $\langle Q^{2} \rangle$.
We use
\begin{equation}
\frac{d\sigma}{dt} 
= \frac{d\sigma_{\rm T}}{dt} + \varepsilon \frac{d\sigma_{\rm L}}{dt}
\label{3.100}
\end{equation}
with $\varepsilon = 1$
($\varepsilon \approx 1$ for the HERA kinematic region).
In the top panels of Fig.~\ref{fig:3}
the upper line corresponds to $W = 12.7$~GeV 
and $Q^{2} = 0$ and should be compared 
to the averaged FNAL data (top data points)
for the $\gamma p \to \gamma p$ reaction
\cite{Breakstone:1981wm}.
In this kinematic range, for $Q^{2} = 0$
and at intermediate $W$, 
the reggeon plus soft-$\Pom$ contributions 
dominate and the hard-$\Pom$ exchange 
gives a negligible contribution.
As expected, there is a significant interference between 
the reggeon and soft-pomeron components.
The slope parameters $b_{2}$ and $b_{1}$ in (\ref{t_dependence_ff})
are adjusted to the FNAL $d\sigma/dt$ data on 
the real-photon-proton scattering.
At higher $W$ and $Q^{2}$ measured at HERA
the hard pomeron plays an increasingly important role.
The slope parameter $b_{0}$ for the hard-pomeron exchange
is adjusted to the DVCS HERA data.
As we noted above, $d\sigma(\gamma^{*} p \to \gamma p)/dt$
is the sum of two contributions
$d\sigma_{\rm T}/dt$ and $d\sigma_{\rm L}/dt$ with the latter term
becoming very small for $|t| \to 0$.
This is understandable since for the $\gamma^{*} p \to \gamma p$
forward scattering only the double-helicity-flip amplitudes
can contribute to $d\sigma_{\rm L}/dt$.
Furthermore, we find that $d\sigma_{\rm T}/dt$ is dominated 
by the $b$-type couplings
and $d\sigma_{\rm L}/dt$ is dominated by the $a$-type couplings.

In the middle panels of Fig.~\ref{fig:3},
we show the complete theoretical result
and individual components contributing to
the cross-section $d\sigma/dt$, see (\ref{3.100}),
for $W = 82$~GeV and $Q^{2} = 8$~GeV$^{2}$
together with the H1 data \cite{H1:2009wnw}.
The contributions of soft $\Pom$ (the blue short-dashed lines), 
hard $\Pom$ (the red long-dashed lines),
the interference term (the green dotted lines),
and their sum total (the thin full lines)
for ${\rm T}$ and ${\rm L}$ components individually are also shown.
The constructive interference of the soft 
and hard pomeron terms is a salient feature there.

In the bottom panels of Fig.~\ref{fig:3} we show 
the ratios of the $\gamma^{*} p \to \gamma p$ cross-sections
for longitudinally and transversely polarized
virtual photons,
\begin{eqnarray}
R(Q^{2},W^{2}) = \frac{\sigma_{\rm L}(Q^{2},W^{2})}{\sigma_{\rm T}(Q^{2},W^{2})} \,, \quad
\tilde{R}(Q^{2},W^{2},t) = \frac{\frac{d\sigma_{\rm L}}{dt}(Q^{2},W^{2},t)}
                                  {\frac{d\sigma_{\rm T}}{dt}(Q^{2},W^{2},t)} \,,
\label{ratio_t}
\end{eqnarray}
as functions of $Q^{2}$ and $|t|$, respectively.
The cross-section $\sigma_{\rm L}$
vanishes proportionally to $Q^{2}$ for $Q^{2} \to 0$.
The ratio $\tilde{R}(Q^{2},W^{2},t)$ strongly grows with $|t|$.
We must emphasize that this behaviour of 
$\tilde{R}(Q^{2},W^{2},t)$
depends crucially on our (reasonable) assumption
that $a$ and $b$ couplings in the $\Pom_{j} \gamma^{*} \gamma^{*}$
vertex functions have the same $t$ dependence 
for a given $j$.

In \cite{Capua:2006ij,Fazio:2013hza} 
also the Regge theory was applied to DVCS. 
These authors consider only leading helicity amplitudes
(only transversely polarized initial $\gamma^{*}$'s).
In contrast, in our tensor-pomeron approach 
we present a complete model (all helicity amplitudes).
Therefore, we could make, e.g., predictions for
$\sigma_{\rm L}/\sigma_{\rm T}$ which can and should be checked
by experiments.
The contribution of the interference term
found in \cite{Fazio:2013hza} is considerable
for intermediate values of $Q^{2}$, 
but smaller than our findings.

\newpage 
\section{Conclusions}

The two-tensor-pomeron model proposed previously 
to describe low $x$ DIS data \cite{Britzger:2019lvc}
was applied to deeply virtual Compton scattering (DVCS)
for high c.m. energies $W$ and small Bjorken $x$, 
say $x \lesssim 0.02$ \cite{Lebiedowicz:2022xgi}.

In particular, the transition from the small-$Q^{2}$ regime,
including the photoproduction ($Q^{2} = 0$) limit,
to the large-$Q^{2}$ regime, the DIS limit is well described.
We compared predictions of the two-tensor-pomeron model 
to the DVCS data measured at HERA.
We onsidered FIT~1 
in which a 'minimal' modification of the $Q^2$ dependence 
of only one $\gamma^*(Q^2) \gamma \Pom$ coupling function
was made.
We considered also FIT~2
in which the size of the hard-pomeron component was increased,
especially for larger $Q^2$,
and the soft-pomeron component was 
reduced relative to FIT~1.
We kept here, on purpose, the same parameters of the form factors (\ref{t_dependence_ff}) as in FIT~1.
The FIT~2 better describes the data 
at larger $|t|$ for $Q^2 \gtrsim 8$~GeV$^{2}$ 
(see Fig.~\ref{fig:3}).

The model describes the $W$, $Q^2$ and $t$ dependences of
$d\sigma(\gamma^{*} p \to \gamma p)/dt$ measured at HERA
and of the elastic photon-proton cross-section measured at FNAL.
A good description of the DVCS data 
is achieved due to a sizeable interference 
of soft and hard pomeron contributions.
The soft component and also the interference of soft and hard terms
are very important up to at least $Q^{2} \simeq 20$~GeV$^{2}$.

Our calculation include the contributions of both 
the transverse and longitudinal virtual photons. 
The longitudinal cross-section $d\sigma_{\rm L}/dt$
is predicted to be very small for $|t| \to 0$ but
to be sizeable for 
$0.5 \; {\rm GeV}^{2} \lesssim |t| \lesssim 1.0\; {\rm GeV}^{2}$.
We give also predictions for
$\sigma_{\rm L}/\sigma_{\rm T}$ which can and should be checked
by experiments.
The corresponding ratio of ${\rm L/T}$ grows strongly with $|t|$.
We showed the $Q^2$ dependence of this ratio 
for different c.m. energies of the $\gamma^{*}p$ system.

We presented predictions for low-$x$ DVCS
of the two-tensor-pomeron model which previously
was successfully applied to low $x$ DIS in \cite{Britzger:2019lvc}.
The model provides amplitudes for all helicity configurations
and, thus, can be checked by experimentalists in many ways.
We are looking forward to further tests of 
the non-perturbative QCD dynamics embodied in our
tensor-pomeron exchanges
in future electron-hadron collisions 
in the low-$x$ regime at the EIC \cite{AbdulKhalek:2021gbh} 
and LHeC \cite{LHeC:2020van} colliders.

\section*{Acknowledgments}
P.L. thanks the organizers of XXIX Cracow Epiphany Conference
for a well-organized conference with many stimulating discussions.
This work was partially supported by
the Polish National Science Centre Grant
No. 2018/31/B/ST2/03537.

\end{document}